\documentclass[12pt,dvips]{article}

\usepackage{rotating}
\usepackage{hhline}
\usepackage{array}
\usepackage{epsfig}

\begin{document}

\renewcommand{\thefootnote}{\fnsymbol{footnote}}
\tolerance=100000

\newcommand{\lsim}{\raisebox{-0.13cm}{~\shortstack{$<$\\[-0.07cm] $\sim$}}~}
\newcommand{\gsim}{\raisebox{-0.13cm}{~\shortstack{$>$\\[-0.07cm] $\sim$}}~}

\newcommand{\imag}{\Im {\rm m}}
\newcommand{\real}{\Re {\rm e}}
\newcommand{\s}{\\ \vspace*{-3.5mm}}

\def\tablename{\bf Table}%
\def\figurename{\bf Figure}%

\begin{flushright}
KIAS--P03079 \\[-0.1cm]
KUPT--03--04 \\[-0.1cm]
hep-ph/0311037\\
\today
\end{flushright}

\vspace{1.2cm}

\begin{center}
  {\Large \bf Analysis of the Neutralino System in Two--Body Decays
              of Neutralinos}\\[1.5cm]
              S. Y. Choi$^1$ and Y. G. Kim$^2$
\end{center}

\vskip 0.5cm

{\small
\begin{enumerate}
\item[{}] $^1$ {\it Department of Physics, Chonbuk National University,
               Chonju 561-756, Korea}\\[-1.cm]
\item[{}] $^2$ {\it Department of Physics, Korea University,
               Seoul 136--701, Korea}
\end{enumerate}
}

\renewcommand{\thefootnote}{\fnsymbol{footnote}}
\vspace{3.cm}

\begin{abstract}
\noindent
In the minimal supersymmetric standard model (MSSM), the neutralinos,
the spin--1/2 Majorana superpartners of the neutral gauge and Higgs bosons,
are expected to be among the light supersymmetric particles that can
be produced copiously at future high--energy colliders.
We analyze two--body neutralino decays into a neutralino plus a $Z$ boson
or a lightest neutral Higgs boson $h$, allowing the relevant parameters to
have complex phases. We show that the two--body tree--level decays of
neutralinos are kinematically allowed in a large region of the MSSM
parameter space and they can provide us with a powerful probe of the
Majorana nature and CP properties of the neutralinos through the $Z$--boson
polarization measured from $Z$--boson leptonic decays.
\end{abstract}

\newpage

\section{Introduction}

The search for supersymmetry (SUSY) is one of the main goals at
present and future colliders \cite{futurecoll} as SUSY is generally
accepted as one of the most promising concepts for physics
beyond the Standard Model (SM) \cite{susy}. A special feature of SUSY
theories is the existence of the neutralinos, the spin--1/2 Majorana
superpartners of the neutral gauge bosons and Higgs bosons.
In the MSSM, the neutralinos are expected to be among the light supersymmetric
particles that can be produced copiously at future high--energy
colliders \cite{snowmass}. Once several neutralino candidates are observed
at such high--energy colliders, it will be crucial to establish the Majorana
nature and CP properties of the neutralinos. In this light, many extensive
studies of the general characteristics of the neutralinos in their production
and decays \cite{ckmz,Gudi,choi,11a} as well as in the selectron pair
production \cite{aguilar} at $e^+e^-$ and/or $e^-e^-$ linear colliders have
been performed. \s

In the present work, we analyze two--body tree--level decays
of neutralinos into a neutralino plus a $Z$ boson or a lightest neutral
Higgs boson $h$ in order to probe the Majorana nature of the neutralinos
and CP violation in the neutralino system. A comprehensive analysis of the
two--body decays of neutralinos as well as charginos was given previously
in Ref.~\cite{haber_gunion}. We however note that a rather light Higgs boson
mass was assumed and no $Z$ boson polarization was considered in the
previous work. One powerful diagnostic tool in the present analysis is $Z$
polarization, which can be reconstructed with great precision through
$Z$--boson leptonic decays, $Z\rightarrow l^+l^-$, in particular, with
$l=e, \mu$.\s

It is possible that due to the masses of the relevant particles, no two--body
tree--level decays are allowed, in which case the dominant decays would consist
of three--body tree--level \cite{nojiri} or two--body one--loop
decays \cite{loop_decay}. However, a sufficiently heavy neutralino can decay
via tree--level two--body channels containing a $Z$ or $h$ with its mass less
than 135 GeV in the context of the MSSM \cite{h_mass_bound}. If some sfermions
are sufficiently light, two--body tree--level decays of neutralinos
into a fermion and a sfermion may be also be important. However, neutralinos
heavier than the squarks will be extremely difficult to isolate at
hadron colliders, because the squarks and gluinos are strongly produced
and they decay subsequently into lighter neutralinos and charginos.
On the other hand, at $e^+e^-$ colliders, squarks and sleptons, if they
are kinematically accessible, are fairly easy to produce and study directly.
With these phenomenological aspects in mind, we assume in the present work that
all the sfermions are heavier than (at least) the second lightest neutralino
$\tilde{\chi}^0_2$. Then, we investigate the MSSM parameter space for
the two--body tree--level decays of the neutralino $\tilde{\chi}^0_2$
and show how the Majorana nature and CP properties of the neutralinos
can be probed through the two--body decays $\tilde{\chi}^0_2\rightarrow
\tilde{\chi}^0_1\,Z$, once such two--body decays are kinematically allowed. \s

The paper is organized as follows. Section~\ref{sec:mixing} is devoted to
a brief description of the mixing for the neutral gauginos and higgsinos in
CP--noninvariant theories with non--vanishing phases.
In Sec.~\ref{sec:two-body}, after explaining the reconstruction of
$Z$--boson polarization through the $Z$ decays into
two--lepton pairs, we present the formal description of the (polarized) decay
widths of the two--body neutralino decays into a lightest neutralino
$\tilde{\chi}^0_1$ plus a $Z$ boson or a lightest Higgs boson $h$ with special
emphasis on the polarization of the $Z$ boson. In Sec.~\ref{sec:analysis},
we first investigate the region of the MSSM parameter space where the two--body
neutralino decays are allowed and discuss the dependence of the branching
ratios and decay widths on the relevant SUSY parameters. Then, we give a
simple numerical demonstration of how the Majorana nature and CP properties
of the neutralinos can be probed through the two--body decays
$\tilde{\chi}^0_2\rightarrow\tilde{\chi}^0_1\,Z$. Finally, we conclude in
Sec.~\ref{sec:conclusion}.

\section{Neutralino Mixing}
\label{sec:mixing}

In the MSSM, the mass matrix of the spin-1/2 partners of the neutral
gauge bosons, $\tilde{B}$ and $\tilde{W}^3$, and of the neutral
Higgs bosons, $\tilde H_1^0$ and $\tilde H_2^0$, takes the form
\begin{eqnarray}
{\cal M}_N=\left(\begin{array}{cccc}
  M_1  &  0   &  -m_Z c_\beta s_W  &  m_Z s_\beta s_W \\[1mm]
   0   & M_2  &   m_Z c_\beta c_W  & -m_Z s_\beta c_W\\[1mm]
-m_Z c_\beta s_W & m_Z c_\beta c_W &  0   & -\mu  \\[1mm]
 m_Z s_\beta s_W &-m_Z s_\beta c_W & -\mu &  0
               \end{array}\right)\,,
\label{eq:massmatrix}
\end{eqnarray}
in the $\{\tilde{B},\tilde{W}^3,\tilde{H}^0_1,\tilde{H}^0_2\}$
basis. Here $M_1$ and $M_2$ are the fundamental SUSY
breaking U(1) and SU(2) gaugino mass parameters, and $\mu$ is the
higgsino mass parameter. As a result of electroweak symmetry
breaking by the vacuum expectation values of the two neutral Higgs
fields $v_1$ and $v_2$ ($s_\beta =\sin\beta$, $c_\beta=\cos\beta$
where $\tan\beta=v_2/v_1$), non--diagonal terms proportional to
the $Z$--boson mass $m_Z$ appear and  the gauginos and higgsinos
mix to form the four neutralino mass eigenstates
$\tilde{\chi}_i^0$ ($i=1$--$4$), ordered according to increasing mass.
In general the mass parameters $M_1$, $M_2$ and $\mu$ in the neutralino
mass matrix (\ref{eq:massmatrix}) can be complex. By re--parameterization
of the fields, $M_2$ can be taken real and positive, while the U(1)
mass parameter $M_1$ is assigned the phase $\Phi_1$ and the
higgsino mass parameter $\mu$ the phase $\Phi_\mu$. For the sake of
our latter discussion, it is worthwhile to note that in the
limit of large $\tan\beta$ the gaugino--higgsino mixing becomes
almost independent of $\tan\beta$ and the neutralino sector
itself becomes independent of the phase $\Phi_\mu$ in this limit.\s

The neutralino mass eigenvalues $m_i\equiv m_{\tilde{\chi}^0_i}$
($i=1$-$4$) can  be chosen positive by a suitable definition of
the mixing matrix $N$, rotating the gauge eigenstate basis
$\{\tilde{B},\tilde{W}^3,\tilde{H}^0_1,\tilde{H}^0_2\}$ to the
mass eigenstate basis of the Majorana fields:
$N^*{\cal M}_N\,N^\dagger = {\rm diag} (m_{\tilde{\chi}^0_1},
m_{\tilde{\chi}^0_2}, m_{\tilde{\chi}^0_3}, m_{\tilde{\chi}^0_4})$.
In general the mixing matrix $N$ involves 6 non--trivial angles
and 9 non--trivial phases, which can be classified into three Majorana
phases and six Dirac phases \cite{ckmz}. The neutralino sector is CP
conserving if $\mu$ and $M_1$ are real, which is equivalent to
vanishing Dirac phases (mod $\pi$) and Majorana phases (mod $\pi/2$).
Majorana phases of $\pm \pi/2$ do not signal CP violation
but merely indicate different intrinsic CP parities of the
neutralino states in CP--invariant theories \cite{11a}.\s

\section{Two--Body Decays of Neutralinos}
\label{sec:two-body}

Before describing the two--body decays $\tilde{\chi}^0_i\rightarrow
\tilde{\chi}^0_j\, Z$ in detail, we explain how to reconstruct the $Z$
polarization through the lepton angular distributions of the $Z$--boson
leptonic decays, $Z \rightarrow l^-l^+$, particularly with $l=e, \mu$.
In the rest frame of the decaying $Z$ boson,
which can be reconstructed with great precision by measuring the
lepton momenta, the lepton angular distributions are given by
\begin{eqnarray}
&& \frac{1}{\Gamma[Z\rightarrow l^+l^-]}\,
   \frac{d\Gamma[Z(\pm)\rightarrow l^+l^-]}{d\cos\theta_l}
   =
   \frac{3}{8}\left[1+\cos^2\theta_l\pm 2\,\xi_l \cos\theta_l\right]\,,
   \nonumber\\
&& \frac{1}{\Gamma[Z\rightarrow l^+l^-]}\,
   \frac{d\Gamma[Z(0)\rightarrow l^+l^-]}{d\cos\theta_l}
   =
   \frac{3}{4}\,\sin^2\theta_l\,,
\label{eq:polar_angle}
\end{eqnarray}
for the $Z$--boson helicities, $\pm 1$ and $0$, respectively, where
$\xi_l= 2v_l a_l/(v^2_l+a^2_l)\simeq -0.147$ with $v_l= s^2_W - 1/4$ and
$a_l=1/4$, and $\theta_l$ is the polar angle of the $l^-$ momentum with
respect to the $Z$ boson polarization direction. Here, the decay width
$\Gamma[\,Z\rightarrow l^+l^-]$ is the average of three polarized decay
widths,
\begin{eqnarray}
\Gamma[\, Z\rightarrow l^+l^-] =\frac{1}{3} \left\{\,
\Gamma[\, Z(+)\rightarrow l^+l^-]+\Gamma[\, Z(0)\rightarrow l^+l^-]
+\Gamma[\, Z(-)\rightarrow l^+l^-]\,\right\}.
\end{eqnarray}
We emphasize that the three polar--angle distributions (\ref{eq:polar_angle})
can be determined without knowing the full kinematics of the decay
$\tilde{\chi}^0_i\rightarrow\tilde{\chi}^0_j\, Z$. In contrast, the
distributions involving the interference of the amplitudes with different
$Z$ helicities are always accompanied with azimuthal angle dependent terms.
As the lightest neutralino $\tilde{\chi}^0_1$ assumed to be the lightest SUSY
particle (LSP) always escapes detection, the kinematics of the two--body decay
$\tilde{\chi}^0_i\rightarrow\tilde{\chi}^0_j\, Z$ cannot be fully reconstructed
so that the azimuthal--angle dependent distributions are not fully available.\s

The decay width of the decay $\tilde{\chi}^0_i\rightarrow\tilde{\chi}^0_j\,
Z$ producing a $Z$ boson with its helicity, $\pm 1$ or $0$, reads
\begin{eqnarray}
&& \Gamma\left[\tilde{\chi}^0_i\rightarrow \tilde{\chi}^0_j\,Z(\pm)\right]
  = \frac{g^2_Z\,\lambda^{1/2}_Z}{16\pi m^3_i}\,(|V|^2+|A|^2)
     \left[\,m^2_i+m^2_j-m^2_Z- 2m_i m_j\, {\cal A}_N
      \pm \frac{\lambda^{1/2}_Z}{2} {\cal A}_T\right]\,,
     \nonumber\\
&& \Gamma\left[\tilde{\chi}^0_i\rightarrow \tilde{\chi}^0_j\, Z(0)\right]
  = \frac{g^2_Z\,\lambda^{1/2}_Z}{16\pi m^3_i}\,(|V|^2+|A|^2)
     \left[\frac{\lambda_Z}{m^2_Z}+m^2_i+m^2_j-m^2_Z
           - 2\,m_i m_j\,{\cal A}_N\,\right]\,,
\label{eq:polarized_decay_width}
\end{eqnarray}
respectively, where the asymmetries $A_N$ and $A_T$ are defined in terms of
the vector and axial--vector couplings $V$ and $A$ of the $Z$ boson to the
neutralino current as
\begin{eqnarray}
{\cal A}_N = \frac{|V|^2-|A|^2}{|V|^2+|A|^2},\qquad
{\cal A}_T = \frac{2\, \real(VA^*)}{|V|^2+|A|^2}\,,
\label{eq:asymmetry}
\end{eqnarray}
with $g_Z=g/\cos\theta_W$ and
the kinematical factor $\lambda_Z =[(m_i+m_j)^2-m^2_Z][(m_i-m_j)^2-m^2_Z]$.
Combining the leptonic $Z$--boson decay distributions (\ref{eq:polar_angle})
with the polarized decay widths (\ref{eq:polarized_decay_width}), we
obtain the correlated polar--angle distribution:
\begin{eqnarray}
\frac{d\Gamma_{\rm corr}}{d\cos\theta_l}
 &=&\frac{3}{8}\,{\cal B}[Z\rightarrow ll]\,\bigg\{
  \left(\Gamma[\,\tilde{\chi}^0_i\rightarrow\tilde{\chi}^0_j\, Z(+)]
       +\Gamma[\,\tilde{\chi}^0_i\rightarrow\tilde{\chi}^0_j\, Z(-)]
  \right)\, (1+\cos^2\theta_l)\nonumber\\
  && { }\hskip 2.cm
 +2\left(\Gamma[\,\tilde{\chi}^0_i\rightarrow\tilde{\chi}^0_j\, Z(+)]
       -\Gamma[\,\tilde{\chi}^0_i\rightarrow\tilde{\chi}^0_j\, Z(-)]
  \right)\, \xi_l\,\cos\theta_l \nonumber\\
  && { }\hskip 2.cm
 + 2\,\Gamma[\,\tilde{\chi}^0_i\rightarrow\tilde{\chi}^0_j\, Z(0)]
    \,\sin^2\theta_l\bigg\}\,.
 \label{eq:correlated_polar_angle}
\end{eqnarray}
Consequently, each polarized decay width can be extracted from the correlated
polar--angle distribution  by projecting out the distribution with a proper
lepton--polar angle distribution.\s

The explicit forms of the vector and axial--vector couplings $V$ and $A$ in
Eq.~(\ref{eq:asymmetry}) are given in terms of the
$4\times 4$ neutralino diagonalization matrix $N$ in the MSSM by
\begin{eqnarray}
V = -\frac{i}{2}\imag\left(N_{j3}N^*_{i3}-N_{j4}N^*_{i4}\right)\,,\qquad
A = \frac{1}{2}\real\left(N_{j3}N^*_{i3}-N_{j4}N^*_{i4}\right)\,.
\label{eq:VA}
\end{eqnarray}
Note that {\it the vector coupling $V$ is pure imaginary and the axial--vector
coupling $A$ is pure real.} This characteristic property of the
$Z$-$\tilde{\chi}^0_i$-$\tilde{\chi}^0_j$ coupling due to the Majorana nature
of neutralinos leads to one important relation between the polarized decay
widths with the $Z$--boson helicities, $\pm 1$:
\begin{eqnarray}
  \Gamma[\,\tilde{\chi}^0_i\rightarrow\tilde{\chi}^0_j Z(+)]
 \, =\, \Gamma[\,\tilde{\chi}^0_i\rightarrow\tilde{\chi}^0_j Z(-)]\,,
\label{eq:Majorana}
\end{eqnarray}
which is valid even in the CP non--invariant theory. This relation can be checked
by measuring the forward--backward polar--angle asymmetry of the correlated
polar--angle distribution (\ref{eq:correlated_polar_angle}). However, because of
the small analyzing power $\xi_l\simeq -0.147$, it will be necessary to have
sufficient large number of decay events to measure the asymmetry with good
precision. In addition to the relation (\ref{eq:Majorana}), the relative intrinsic
CP parity of two neutralinos in the CP invariant theory can be determined by
measuring the ratio of the longitudinal decay width to the transverse decay width,
which satisfies
\begin{eqnarray}
{\cal R}_{LT}\equiv
   \frac{2\Gamma[\,\tilde{\chi}^0_i\rightarrow\tilde{\chi}^0_j Z(0)]}{
   \Gamma[\,\tilde{\chi}^0_i\rightarrow\tilde{\chi}^0_j Z(+)]
+  \Gamma[\,\tilde{\chi}^0_i\rightarrow\tilde{\chi}^0_j Z(-)]}
   = \frac{(m_i\mp m_j)^2}{m^2_Z}\,,
\label{eq:ratio R}
\end{eqnarray}
for the even/odd relative intrinsic CP parity with $V=0$/$A=0$, {\it i.e.}
${\cal A}_N=\mp 1$, respectively. In the CP non--invariant theory, both the
vector and axial couplings are in general non--vanishing, leading to the
value of the asymmetry $A_N$ different from $\pm 1$.
Therefore, any precise measurements of the asymmetry $A_N$ will provide
us with an important probe of CP violation in the neutralino system under
the assumption that {\it the neutralino masses are measured with good
precision, independently of the decay modes.}\s

Next, we give the decay formulas into final states containing a lightest
neutral Higgs boson $h$. The explicit form of the decay width of the decay
$\tilde{\chi}^0_i\rightarrow \tilde{\chi}^0_j\, h$ is written as
\begin{eqnarray}
\Gamma\left[\tilde{\chi}^0_i\rightarrow \tilde{\chi}^0_j\,h\right]
  = \frac{g^2\,\lambda^{1/2}_h}{16\pi m^3_i}
     \left[|S|^2\,((m_i+m_j)^2-m^2_h) + |P|^2\, ((m_i-m_j)^2-m^2_h)\right]\,,
\end{eqnarray}
with the kinematical factor $
\lambda_h =[(m_i+m_j)^2-m^2_h][(m_i-m_j)^2-m^2_h]$.
The scalar and pseudoscalar couplings, $S$ and $P$, of the Higgs boson $h$
to the neutralino current are defined in terms of the mixing matrix $N$ as
\begin{eqnarray}
&& S = \frac{1}{2}\,\real\left[(N_{j2}-t_W N_{j1})(s_\alpha N_{i3}+c_\alpha N_{i4})
                           +(i\leftrightarrow j)\right]\,,\nonumber\\
&& P = \frac{i}{2}\imag\left[(N_{j2}-t_W N_{j1})(s_\alpha N_{i3}+c_\alpha N_{i4})
                           +(i\leftrightarrow j)\right]\,,
\end{eqnarray}
where $t_W=\tan\theta_W$, $c_\alpha=\cos\alpha$ and $s_\alpha=\sin\alpha$ for the
neutral Higgs mixing angle $\alpha$. If the charged Higgs boson mass in the MSSM is
very large,
\begin{eqnarray}
c_\alpha \rightarrow \sin\beta,\qquad
s_\alpha \rightarrow -\cos\beta\,,
\end{eqnarray}
This decoupling approximation of the cosine and sine of the mixing angle $\alpha$ is
very good if the charged Higgs mass is larger than twice the $Z$ boson
mass \cite{decoupling}. For the sake of discussion, we take
the decoupling limit in the present work.\s

\section{Numerical Analysis of Two--Body Decays}
\label{sec:analysis}

In some SUSY scenarios, the lightest neutralino $\tilde{\chi}^0_1$ is the
LSP and the second lightest neutralino $\tilde{\chi}^0_2$ among the other
three neutralino states are expected to be lighter than sfermions and
gluino \cite{snowmass}. Then, the two--body decays $\tilde{\chi}^0_2
\rightarrow\tilde{\chi}^0_1\, Z$ or $h$ as well as the two--body decays of
the heavier neutralinos $\tilde{\chi}^0_{3,4}$ \cite{haber_gunion} will
constitute the major decay modes of the neutralinos, respectively, once
the two--body tree--level decay modes are kinematically allowed. In the
following numerical analysis we will ignore all other modes except for
the two--body tree--level decays of the neutralinos.

\subsection{Branching ratios}

For the branching ratio calculations for the two--body decays
$\tilde{\chi}^0_2\rightarrow\tilde{\chi}^0_1\,Z/h$, we assume that all the SUSY
parameters are real, $M_1$ is related to $M_2$ by the gaugino mass unification
condition $|M_1|=(5/3)\,\tan^2\theta_W M_2 \approx 0.5\, M_2$ and the Higgs
boson mass $m_h$ is 115 GeV. In addition, we assume that the MSSM Higgs system
is in the decoupling regime so that the characteristics of the lightest Higgs
boson $h$ is similar to the SM Higgs boson to a good
approximation \cite{decoupling}.\s

\begin{figure}[htb]
\begin{center}
\mbox{ }\\[-1.cm]
\epsfig{file=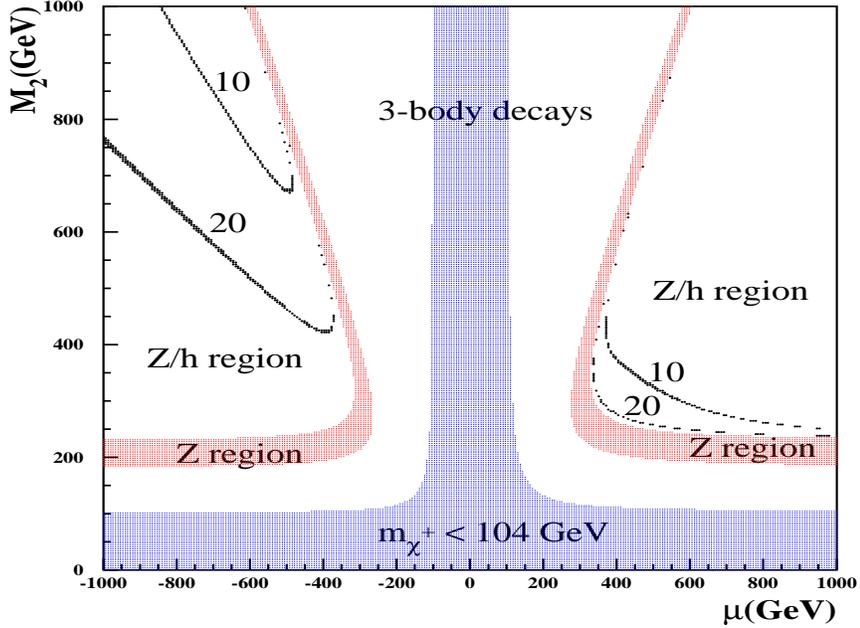,height=10cm,width=13cm}
\vskip -0.5cm
\caption{\it Three distinct regions of $\tilde{\chi}^0_2$ decay are exhibited as
             a function of $M_2$ and $\mu$, assuming that $M_2$ and $\mu$ are
             real. In the region denoted by ``three--body decays", no two--body
             modes for the neutralino $\tilde{\chi}^0_2$ except for
             the loop--induced two--body radiative decays are kinematically
             allowed. In the ``$Z$ region" (red--colored), only the
             two--body decay into a $Z$ is allowed and in the ``$Z/h$ region",
             both the two--body decays are allowed. The ``$Z/h$ region" is
             divided into three parts, according to ${\cal B}[Z]\leq 10\%$,
             $10\% \leq {\cal B}[Z]\leq 20\%$ and ${\cal B}[Z]\geq 20\%$.
             For reference, the exclusion region by the experimental bound
             on the lighter chargino mass bound
             $ m_{\tilde{\chi}_1^\pm }\geq 104$ GeV is displayed by the
             blue--hatched region. In this numerical illustration, we set
             $\tan\beta=10$.}
\label{fig:bratio}
\end{center}
\end{figure}

Figure~\ref{fig:bratio} shows the regions of the decays of
$\tilde{\chi}^0_2$ on the $\{\mu, M_2\}$ plane. In the region denoted by
``three--body decays", no two--body modes are kinematically allowed. In the
``$Z$ region" (red--colored), only the two--body decay into a $Z$ is allowed
and in the ``$Z/h$ region", both the two--body decays
$\tilde{\chi}^0_2\rightarrow\tilde{\chi}^0_1\, Z/h$ are allowed. We divide
the ``$Z/h$ region" into three parts, according to ${\cal B}[Z]\leq 10\%$,
$10\% \leq {\cal B}[Z]\leq 20\%$ and ${\cal B}[Z]\geq 20\%$. (\,Here,
${\cal B}[Z]\equiv {\rm Br}[\tilde{\chi}^0_2\rightarrow\tilde{\chi}^0_1\,Z]$.)
In addition, as a reference, the region excluded by the experimental
bound \cite{PDG} on the lighter chargino mass $m_{\tilde{\chi}_1^\pm}\geq 104$
GeV is displayed by the blue--hatched region.\s

We first note that, if $M_2\lsim 2m_Z$, the mass difference
$m_{\tilde{\chi}^0_2}-m_{\tilde{\chi}^0_1}$ is less than $m_Z$
for all $\mu$ and the mass difference is very small for $|\mu|\ll M_1, M_2$.
So, as clearly shown in Fig.~\ref{fig:bratio} the two--body decay
$\tilde{\chi}^0_2 \rightarrow \tilde{\chi}^0_1\, Z$ is allowed only when
$2\, m_Z \lsim  M_2 \lsim 2 |\mu|$ under the assumption of the gaugino mass
unification condition.
In addition, we find from the figure that for the two--body decays the
magnitude of $\mu$ is required to be larger than about 270 GeV and that,
once the two--body Higgs mode $\tilde{\chi}^0_2\rightarrow \tilde{\chi}^0_1 h$
is open kinematically, this two--body decay mode dominates in most of the
$Z/h$ region. The region where the decay $\tilde{\chi}^0_2 \rightarrow
\tilde{\chi}^0_1\, Z$ is appreciable is not symmetric between positive and
negative $\mu$ in the $Z/h$ region. The branching ratio ${\cal B}[Z]$ is
significant only in a small area of the positive $\mu$ region, but in a large
area of the negative $\mu$ region.\s

On the other hand, we find numerically that, for the heavier neutralinos
$\tilde{\chi}^0_{3,4}$, the $\{M_2, \mu\}$ region for the two--body decays
$\tilde{\chi}^0_{3,4}\rightarrow \tilde{\chi}^0_1\, Z/h$ expands drastically.
A large region with small $|\mu|$ but large $M_2$ as well as with small $M_2$
but large $|\mu|$ also allows for the two--body decays of the heavier
neutralinos, $\tilde{\chi}^0_{3,4}\rightarrow \tilde{\chi}^0_1\, Z/h$.
Only in the wedge--shaped band region of the width of about 100 GeV around the
line satisfying the relation $M_2\approx 2 |\mu|$ no two--body decays for the
heavier neutralino $\tilde{\chi}^0_3$ are allowed, while the heaviest neutralino
can still decay into $\tilde{\chi}^0_1$ and $Z/h$ in the (almost) entire
parameter space, possibly except for the region excluded by the experimental
lighter chargino mass bound.\s

Consequently, for most of the parameter space of the MSSM  the decays of the
two heavier neutralinos are dominated by two--body tree--level processes of
which the final state consists of a $Z$ boson or $h$ boson
together with one of the lighter
neutralinos, or a $W$ boson and one of the charginos. Furthermore, the two--body
decays of the second lightest neutralino $\tilde{\chi}^0_2$ can be significant
in a large region of the parameter space of the MSSM.\s

\begin{figure}[htb]
\begin{center}
\epsfig{file=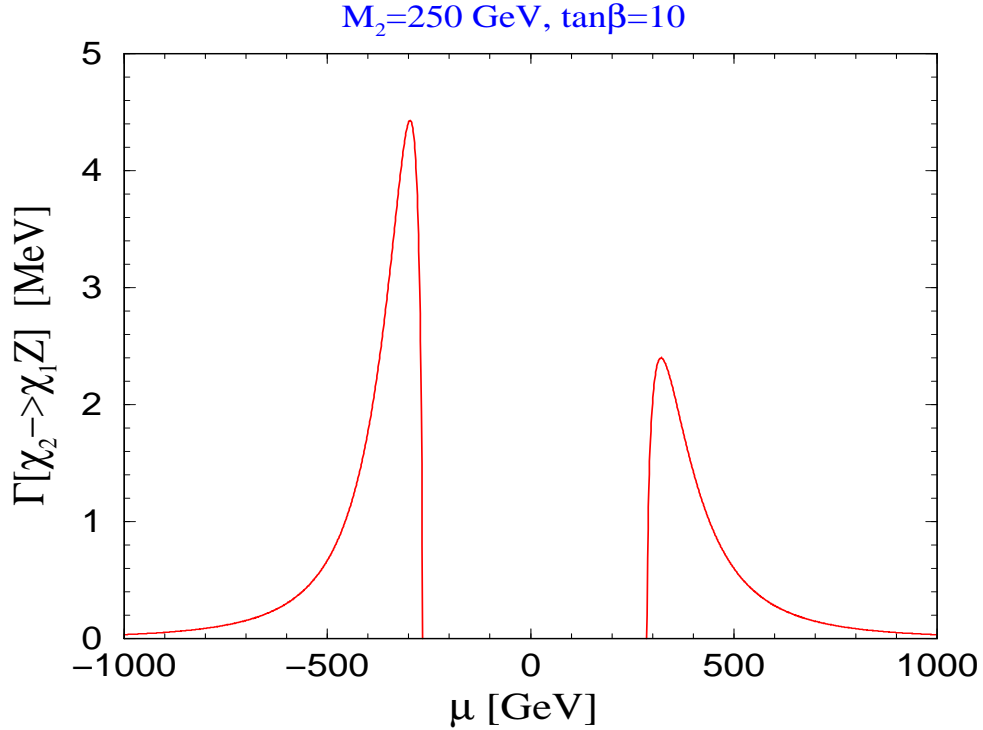,height=10cm,width=13cm}
\caption{\it The dependence of the decay width $\Gamma[\,\tilde{\chi}^0_2
             \rightarrow\tilde{\chi}^0_1\, Z]$ on the higgsino mass parameter
             $\mu$, assuming that $\mu$ is real and $M_1=(5/3)\, t^2_W\, M_2$.
             For this numerical illustration, we set $M_2=250$ GeV and
             $\tan\beta=10$.}
\label{fig:width}
\end{center}
\vskip -0.3cm
\end{figure}

In addition to the branching ratios, it is also crucial to analyze the absolute
size of the decay width $\Gamma[\, \tilde{\chi}^0_2\rightarrow \tilde{\chi}^0_1\, Z]$.
Depending on the values of the relevant couplings, the two--body decay widths could be
smaller than the three--body decay widths involving virtual sfermion exchanges, unless
the sfermions are too heavy.
We exhibit in Fig. \ref{fig:width} the dependence of the decay width
$\Gamma[\,\tilde{\chi}^0_2\rightarrow\tilde{\chi}^0_1\, Z]$ on the higgsino mass parameter
$\mu$, assuming again that $\mu$ is real and taking $M_1=(5/3)\, \tan^2\theta_W\, M_2$,
$M_2=250$ GeV and $\tan\beta=10$. The decay width decreases rapidly with increasing
$|\mu|$. This is because the couplings of the $Z$ boson to the neutralino current are
governed by the higgsino components of the neutralinos (see Eq.~(\ref{eq:VA})) so
that the $Z$-$\tilde{\chi}^0_i$-$\tilde{\chi}^0_j$ couplings are strongly suppressed
for large $|\mu|$. Therefore, for large $|\mu|$, some three--body decays could be more
dominant than the two--body decays. \s

\begin{figure}[htb]
\begin{center}
\epsfig{file=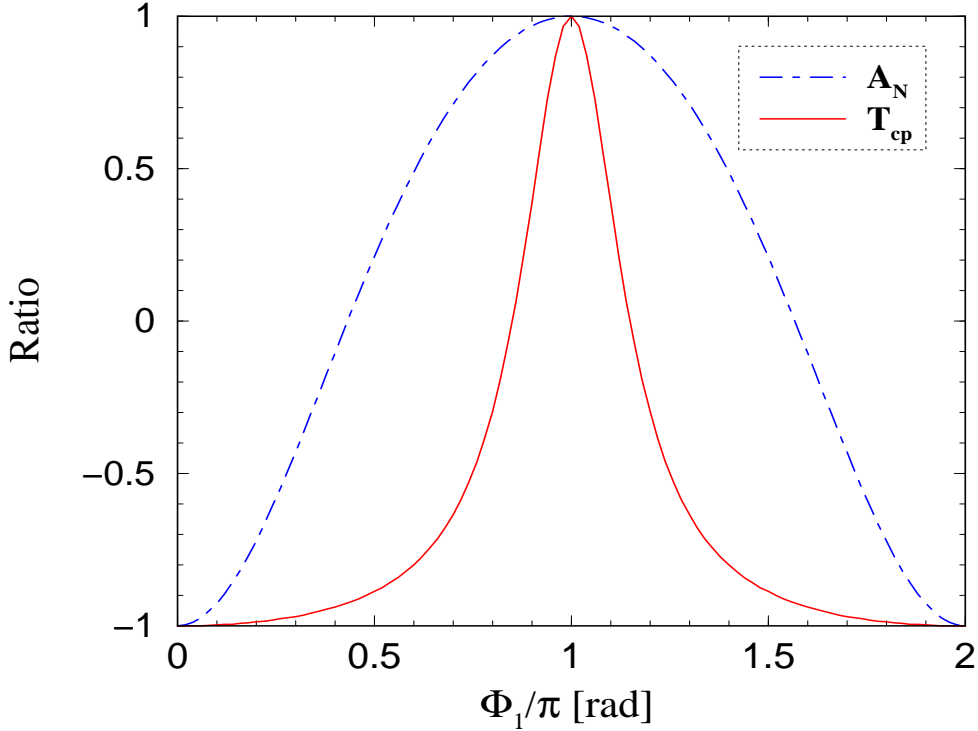,height=10cm,width=13cm}
\caption{\it The dependence of the ratio ${\cal T}_{_{\rm CP}}$ (red solid line) and
             the asymmetry ${\cal A}_N$ (blue dot--dashed line) on the CP phase $\Phi_1$
             for the set of real parameters, $\{\tan\beta=10, M_2= 250\, {\rm GeV},
             |\mu|=500\, {\rm GeV}\}$ . The phase $\Phi_\mu$
             is set zero in this numerical illustration figure, as the ratio
             ${\cal T}_{_{\rm CP}}$ and the asymmetry ${\cal A}_N$ are found to be
             insensitive to the phase $\Phi_\mu$ for the given specific set of real
             parameters.}
\label{fig:etacp}
\end{center}
\vskip -0.3cm
\end{figure}
%

\subsection{A probe of CP violation}

In the previous subsection, we restrict ourselves to the CP invariant case with
real parameters, as the qualitative results obtained from the CP--even quantities
are not expected to change so significantly even if the parameters are complex.
But, the parameters $M_1$ and $\mu$ are in general complex so that it is
important to check whether they indeed have complex phases or not. The existence
of the complex phases in the neutralino system, which in general cause CP
violation, can be established by the measurements of the ratio
\begin{eqnarray}
{\cal T}_{_{\rm CP}}
  =
\frac{{\cal R}_{LT} - (m^2_i+m^2_j)/m^2_Z}{2\, m_i m_j/m^2_Z}\,,
\end{eqnarray}
with ${\cal R}_{LT}$ defined in Eq.~(\ref{eq:ratio R}) as well as
the neutralino masses $m_i$ and $m_j$. The ratio ${\cal T}_{_{\rm CP}}$
is $-1$ ($+1$)  in the CP invariant theory for the positive (negative) relative
intrinsic CP parity of the neutralinos, $\tilde{\chi}^0_i$ and $\tilde{\chi}^0_j$,
taking part in the decay $\tilde{\chi}^0_i\rightarrow \tilde{\chi}^0_j\, Z$,
respectively.\s

To explicitly show the dependence of the ratio ${\cal T}_{_{\rm
CP}}$ on the CP phases $\Phi_1$ and $\Phi_\mu$, we chose a
specific set of real parameters $\{\tan\beta=10, M_2= 250\, {\rm
GeV}, |\mu|=500\, {\rm GeV}\}$ as a simple numerical example with
$|M_1|=(5/3)\,\tan^2\theta_W\, M_2$, while varying the phases
$\Phi_1$ and $\Phi_\mu$. Numerically, we find that the ratio
${\cal T}_{_{\rm CP}}$ is insensitive to the phase $\Phi_\mu$.
This is because the phase dependence is always accompanied with
$\sin2\beta=2\tan\beta/(1+\tan^2\beta)\approx 0.2$ for
$\tan\beta=10$, which is already small, and the higgsino
components of the neutralinos are small for large $|\mu|$. So, we
show in Fig.~\ref{fig:etacp} the dependence of the ratio ${\cal
T}_{_{\rm CP}}$ (red solid) as well as the asymmetry ${\cal A}_N$
(blue dot--dashed) only on the phase $\Phi_1$ for the real
parameter set for one fixed value of $\Phi_\mu=0$. Clearly, in the
CP--invariant case with $\Phi_1=0, \pi$ or $2\pi$, the absolute
magnitude of the ratio ${\cal T}_{_{\rm CP}}$ as well as the
asymmetry ${\cal A}_N$ is 1, but it is different from 1 in the CP
non--invariant case. In the given real parameter set, we find that
the ratio ${\cal T}_{_{\rm CP}}$ is quite sensitive to the phase
$\Phi_1$ near $\Phi_1=\pi$, while it is not so sensitive to the
phase near $\Phi_1=0$ and $2\pi$. \s

In the present work the analysis for probing the Majorana nature
and CP violation in the neutralino system has been carried out at
the tree level. However, it will be important to include loop
corrections to the two--body tree--level decays because, if they
are small, the tree--level CP violation effects might be diluted
by loop--induced CP violation effects originating from other
sectors of the MSSM.\s

\section{Conclusions}
\label{sec:conclusion}

For a large portion of the MSSM parameter space, the decay of the second
lightest neutralino $\tilde{\chi}^0_2$ as well as the heavier neutralinos
$\tilde{\chi}^0_{3,4}$ could be dominated by two--body processes in which
the final state consists of a $Z$ or a lightest Higgs boson $h$ together
with a lightest neutralino $\tilde{\chi}^0_1$, assumed to be the lightest
supersymmetric particle. The main conclusion of the present work is that,
unless the two--body decay $\tilde{\chi}^0_i\rightarrow \tilde{\chi}^0_j\, Z$
is strongly suppressed, the $Z$ polarization, which can be reconstructed
through great precision via the leptonic $Z$--boson decays $Z\rightarrow
l^+l^-$, provides us with a powerful probe of the Majorana nature of the
neutralinos and CP violation in the neutralino system.

\subsection*{Acknowledgments}

The work of SYC was supported in part by the
Korea Research Foundation Grant (KRF--2002--070--C00022) and in
part by KOSEF through CHEP at Kyungpook National University
and the work of YGK was supported by the Korean Federation of Science
and Technology Societies through the Brain Pool program.


\begin{thebibliography}{99}

\bibitem{futurecoll} TESLA Technical Design Report, Part: III Physics at
   an $e^+e^-$ Linear Collider, {\it eds.}\ R.-D. Heuer, D. Miller, F. Richard
   and P. Zerwas, DESY 2001-011 [hep-ph/0106315];
   T. Abe {\it et al.}  [American Linear Collider Working Group Collaboration],
   ``Linear collider physics resource book for Snowmass 2001. 2: Higgs
   and  supersymmetry studies'', in {\it Proc. of the APS/DPF/DPB Summer Study
   on the Future of Particle Physics (Snowmass 2001) }, ed. N. Graf,
   hep-ex/0106056; K. Abe {\it et al.}, JLC Roadmap Report, presented at
   the ACFA LC Symposium,  Tsukuba, Japan 2003, http://lcdev.kek.jp/RMdraft/;
   G. Guignard (ed.), {\it A 3--TeV $e^+e^-$ linear collider based on CLIC
   technology}, CERN--2000-008.

\bibitem{susy} H.P. Nilles, Phys. Rept. {\bf 110} (1984) 110; H. Haber and
   G. Kane, Phys. Rept. {\bf 117} (1985) 75.

\bibitem{snowmass} See, for instance, B.C. Allanach {\it et al.}, Eur. Phys. J.
   C {\bf 25} (2002) 113 [hep--ph/0202233].

\bibitem{ckmz} S.Y. Choi, J. Kalinowski, G. Moortgat-Pick and
   P.M. Zerwas, Eur. Phys. J. C {\bf 22} 563 (2001) [hep-ph/0108117],
   {\it ibid.} C {\bf 23} 769 (2002) [hep-ph/0202039]; J. Kalinowski,
   Acta Phys. Polon. {\bf B34} (2003) 3441 [hep-ph/0306272]; S.Y. Choi,
   hep--ph/0308060.

\bibitem{Gudi} S.T. Petcov, Phys. Lett. {\bf B139} (1984) 421; S.M. Bilenky,
   E.Kh. Khristova and N.P. Nedelcheva. Bulg. J. Phys. {\bf 13} (1986) 283;
   G. Moortgat-Pick and H. Fraas, Phys. Rev. D {\bf 59} (1999) 015016
   [hep-ph/9708481]; G. Moortgat-Pick, H. Fraas, A. Bartl and W. Majerotto,
   Eur. Phys. J. C {\bf 9} (1999) 521 [Erratum-ibid.\ C {\bf 9} (1999) 549]
   [hep-ph/9903220]; G. Moortgat-Pick, A. Bartl, H. Fraas and W. Majerotto,
   Eur. Phys. J. C {\bf 18} (2000) 379 [hep-ph/0007222]; M.M. Nojiri, D. Toya
   and T. Kobayashi, Phys. Rev. D {\bf 62} (2000) 075009 [hep-ph/0001267].

\bibitem{choi} S.M. Bilenky, N.D. Nedecheva and S.T. Petcov, Nucl. Phys.
   {\bf B247} (1984) 61; S.T. Petcov, Phys. Lett. {\bf B178} (1986) 57;
   N. Oshimo, Z. Phys. {\bf C21} (1988) 129; Y. Kizukuri and N. Oshimo, Phys.
   Lett. {\bf B249} (1990) 449; S.Y. Choi, H.S. Song and W.Y. Song, Phys. Rev.
   D {\bf 61} (2000) 075004 [hep-ph/9907474]; V.D. Barger, T. Falk, T. Han,
   J. Jiang, T. Li and T. Plehn, Phys. Rev. D {\bf 64} (2001) 056007
   [hep--ph/0101106]; A. Bartl, H. Fraas, O. Kittel and W. Majerotto,
   hep--ph/0308141; A. Bartl, T. Kernreiter and O. Kittel, hep--ph/0309340;
   S.Y. Choi, M. Drees, B. Gaissmaier and J. Song, hep--ph/0310284.

\bibitem{11a} J. Ellis, J.M. Fr\`ere, J.S. Hagelin, G.L. Kane and S.T. Petcov,
   Phys. Lett. B {\bf 132} (1983) 436; A. Bartl, H. Fraas and W. Majerotto,
   Nucl. Phys. {\bf B278} (1986)1; G. Moortgat-Pick and H. Fraas, Eur. Phys.
   J. C {\bf 25} (2002) 189 [hep-ph/0204333].

\bibitem{aguilar} M. Peskin, in {\it Proc. of the Third Workshop on Physics
   and Experiments with Linear Colliders}, ed. A. Miyamoto {\it et al.},
   (World Scientific, Singapore, 1996) p.248; Int. J. Mod. Phys. {\bf C13}
   (1998) 2299 [hep-ph/9803279]; S. Thomas, Int. J. Mod. Phys. {\bf C13} (1998)
   2307 [hep-ph/9803420];  J.L. Feng and M.E. Peskin, Phys. Rev. D {\bf 64}
   (2001) 115002 [hep--ph/0105100]; A. Freitas, D.J. Miller and P.M. Zerwas,
   Eur. Phys. J. {\bf C21} (2001) 361 [hep--ph/0106198]; A. Datta, A. Djouadi
   and M. Muhlleitner, Eur. Phys. J. {\bf C25} (2002) 539 [hep--ph/0204354];
   C. Blochinger, H. Fraas, G. Moortgat--Pick and W. Porod, Eur. Phys. J.
   {\bf C24} (2002) 297 [hep--ph/0201282]; J.A. Aguilar--Saavedra and A.M.
   Teixeira, hep--ph/0307001; A. Freitas, A. von Manteuffel and P.M. Zerwas,
   hep--ph/0310182.

\bibitem{haber_gunion} J.F. Gunion and H.E. Haber, Phys. Rev. D {\bf 37} (1988)
   2515.

\bibitem{nojiri} M.M. Nojiri and Y. Yamada, Phys. Rev. D {\bf 60} (1999) 015006
   [hep-ph/9902201]; A. Bartl, W. Majerotto and W. Porod, Phys. Lett.
   {\bf B465} (1999) 187 [hep-ph/9907377]; A. Djouadi, Y. Mambrini and
   M. Muhlleitner, Eur. Phys. J. {\bf C20} (2001) 563 [hep--ph/0104115].

\bibitem{loop_decay} H. Komatsu and J. Kubo, Phys. Lett. {\bf B157} (1985) 90;
   H.E. Haber and D. Wyler, Nucl. Phys. {\bf B323} (1989) 267; S. Ambrosanio
   and B. Mele, Phys. Rev. D {\bf 53} (1996) 2541 [hep-ph/9508237]; H. Baer
   and T. Krupovnickas, JHEP {\bf 0209} (2002) 38 [hep--ph/0208277].

\bibitem{h_mass_bound} M.S. Berger, Phys. Rev. D {\bf 41} (1990) 225; Y. Okada,
   M. Yamaguchi and T. Yanagida, Prog. Theor. Phys. {\bf 85} (1991) 1;
   Phys. Lett. {\bf B262} (1991) 54; J. Ellis, G. Ridolfi and F. Zwirner,
   Phys. Lett. {\bf B257} (1991) 83; H.E. Haber and R. Hempfling,
   Phys. Rev. Lett. {\bf 66} (1991) 1815.

\bibitem{decoupling} H.E. Haber and Y. Nir, Phys. Lett. {\bf B306} (1993) 327;
   H.E. Haber, in {\it Physics From the Planck Scale to the Electroweak Scale},
   Proc. of the US--Polish Workshop, Warsaw, Poland, September 21--24, 1994,
   edited by P. Nath, T. Taylor and S. Pokorski (World Scientific, Singapore,
   1995) pp 49--63; J.F. Gunion and H.E. Haber, Phys. Rev. D {\bf 67} (2003)
   075019 [hep--ph/0207010].

\bibitem{PDG} Particle Data Group, K. Hagiwara {\it et al.}, Phys. Rev. D
   {\bf 66} (2002) 01001.

\end{thebibliography}
\end{document}